\documentclass[12pt]{article}

\usepackage{amssymb}
\usepackage{amsmath}
\usepackage{amscd}
\usepackage{latexsym}
\usepackage{graphicx}
\usepackage{booktabs}

\usepackage{cite}

\newcommand{\be}{\begin{equation}}
\newcommand{\ee}{\end{equation}}
\newcommand{\Dlt}{\Delta}

\newcommand{\prt}{\partial}

\newcommand{\bt}{\beta}

\newcommand{\ep}{\varepsilon}

\newcommand{\ra}{\rightarrow}

\newcommand{\gm}{\gamma}

\newcommand{\Gm}{\Gamma}

\newcommand{\lbd}{\lambda}

\begin{document}

\begin{center}

{\Large{\bf Optimized Self-Similar Borel Summation} \\ [5mm]

Simon Gluzman$^1$ and Vyacheslav I. Yukalov$^{2,3}$ } \\ [3mm]

{\it 
$^{1}$Materialica + Research Group \\
Bathurst St. 3000, Apt. 606, Toronto, ON M6B 3B4, Canada\\ [2mm]

$^{2}$Bogolubov Laboratory of Theoretical Physics \\
Joint Institute for Nuclear Research, Dubna 141980 Russia \\ [2mm]

$^{3}$Instituto de Fisica de S\~ao Carlos \\
Universidade de S\~ao Paulo, CP 369, S\~ao Carlos 13560-970, Brazil } \\ [3mm]

{\bf E-mails}: {\it simongluzmannew@gmail.com}, ~~ {\it yukalov@theor.jinr.ru} 

\end{center}

\vskip 2cm

\begin{abstract}
The method of Fractional Borel Summation is suggested in conjunction with self-similar 
factor approximants. The method used for extrapolating asymptotic expansions at small 
variables to large variables, including the variables tending to infinity, is described. 
The method is based on the combination of optimized perturbation theory, self-similar 
approximation theory, and Borel-type transformations. General Borel Fractional 
transformation of the original series is employed. The transformed series is resummed 
in order to adhere to the asymptotic power laws. The starting point is the formulation 
of dynamics in the approximations space by employing the notion of self-similarity. 
The flow in the approximation space is controlled, and ``deep'' control is incorporated 
into the definitions of the self-similar approximants. The class of self-similar 
approximations, satisfying, by design, the power law behavior, such as the use of 
self-similar factor approximants, is chosen for the reasons of transparency, explicitness, 
and convenience. A detailed comparison of different methods is performed on a rather 
large set of examples, employing self-similar factor approximants, self-similar iterated 
root approximants, as well as the approximation technique of self-similarly modified 
Pad\'e - Borel approximations.
\end{abstract}

\vskip 1cm
{\parindent=0pt
{\bf Keywords}:
optimized perturbation theory; extrapolation of asymptotic series; fractional Borel-type 
transforms; factor approximants; critical amplitudes }

\newpage

\section{Introduction}

The problem of extrapolating asymptotic series derived for small variables to finite 
and even very large values of variables is constantly met in various branches of science, 
such as physics, chemistry, economics, applied mathematics, etc. Different approaches 
have been suggested to cure this problem , e.g., Pad\'{e} summation~\cite{bak}, Borel 
summation~\cite{Borel_2}, Shanks transformation \cite{Shanks_3}, hypergeometric Meijer 
approximants~\cite{Shalaby_4,Shalaby_5}, the renormalization group~\cite{Wilson_6,Dupuis_7}, 
and others~\cite{Hardy,beno, Kleinert_10,Kardar_11,L1,L2,L3}. These techniques can provide 
good approximations for finite variables, although they are not applicable for the case 
of variables tending to infinity.   

In the present paper, we describe an original approach based on the transparent 
physical notions of optimization and self-similarity combined with Borel-type summation. 
The layout of the paper is as follows. In Section~\ref{sec2}, we explain the main ideas: 
(i) how to make a divergent asymptotic series convergent and how the convergence is 
optimized by introducing control functions; (ii) how to transform a sequence of approximants 
into a dynamical system leading to the property of self-similarity; (iii) that the solutions 
of the self-similar equation of motion extrapolate the given asymptotic series serving
as a boundary condition into self-similar approximants that are valid for arbitrary values 
of variables. In Section~\ref{sec3}, we combine the self-similar approximants with Borel-type 
summation and demonstrate the ways of introducing control functions or control parameters, 
defined by the minimal difference and minimal derivative optimization conditions. Several 
admissible Borel-type transforms are discussed. In Section~\ref{bas}, we discuss the 
specifics of the critical amplitude calculations. In Section~\ref{U}, we illustrate the 
discussed methods by concrete examples. Section~\ref{conc} concludes the paper. 

In our approach to the resummation problem, we are guided by the requirement for the 
independence of observables from the approximations, transformations, parameters, etc., 
introduced in the course of the analysis. Nature should not be aware of our difficulties 
of understanding it. In turn, we should be respectful and introduce minimal necessary 
assumptions about it. Since the description of the sought function is available to us 
only by means of truncated series, we need to compensate for the lack of knowledge of 
the true coefficients by adding some natural assumptions.

First, the initial truncated series is to be transformed so that, instead of the 
available divergent truncated series, a supposedly better-behaving truncated series 
is considered. Second, to the transformed series, we apply the so-called approximations 
with power law behavior at infinity. At this step, we also reconstruct the behavior 
of the series. Third, the guiding principle of self-similarity leads to self-similar 
roots and self-similar factor approximants. Fourth, from a technical viewpoint, it is 
enormously helpful for the expressions for the critical amplitudes to become explicit 
so that they can be factored into the parts arising from the approximants and from the 
inverse transformation. Fifth, after an optimization with the minimal difference and 
(or) minimal derivative conditions, we study the numerical convergence of the sequences 
of approximations for the sought amplitudes. Besides the numerical convergence, we 
are also guided, when appropriate, by the rigorous results for the convergence of the 
diagonal Pad\'{e} approximants obtained by Gonchar.


\section{Optimization and Self-Similarity}\label{sec2}

In this section, we present the main ideas of the approach in order for the reader to 
grasp the basis of the particular techniques that are used. The two pivotal points
are the notions of convergence optimization and self-similarity.

\subsection{Asymptotic Series}

The starting point of the consideration concerns the well-known fact that, in practical 
applications, we often satisfy the necessity of solving problems by applying some kind of 
perturbation theory, resulting in expansions to powers of a parameter or a variable. 
Such expansions usually represent asymptotic series diverging for finite values of the 
expansion~variable. 

To be specific, let us be interested in a real function $f(x)$ of a real positive 
variable $x$, defined by rather complicated equations that, because of their complexity, 
can be solved only by means of a kind of perturbation theory. The latter would  result 
in an expansion in powers of asymptotically small $x \ra 0$:
\be
\label{1}
 f(x) \simeq f_k(x) \qquad ( x \ra 0 ) ,
\ee
having the form of a truncated series
\be
\label{2}
 f_k(x) = f_0(x) \sum_{n=0}^k a_n x^n ,
\ee
where $f_0(x)$ is a known function. Our main concern is the summation of the power 
expansion, because of which we shall concentrate on the expansion of the form
\be
\label{3}
 f_k(x) =  \sum_{n=0}^k a_n x^n .
\ee
As is evident, the more general form (\ref{2}) can be easily reduced to (\ref{3}). 

In practically all interesting problems, expression (\ref{3}) represents an asymptotic
series diverging for a finite value of $x$. At the same time, the quantities of interest 
usually correspond to finite and sometimes to quite large values of the variable $x$. 
Moreover, in many cases, the most interesting region is the region of very large 
$x \ra \infty$, where the sought function behaves according to the power law
\be
\label{4}
  f(x) \simeq B x^\bt \qquad ( x \ra \infty) .
\ee
Thus, the principal problem is how it would be possible to reconstruct the 
large-variable behavior of the sought function (\ref{4}), knowing only the small-variable 
expansion (\ref{3})?

\subsection{Control Functions}
\label{cf}

It seems clear that, in order to define an effective sum of the divergent series (\ref{3}), 
it is necessary to rearrange the divergent series into a convergent series. When one is
interested in rather large values of $x$, then one cannot resort to Pad\'{e} approximants,
since for asymptotically large $x$ values, the limit of a Pad\'{e} approximant is proportional
to $x^{M-N}$, depending on which of the approximants from the Pad\'{e} table is accepted.   
In that sense, the large-variable behavior of Pad\'{e} approximants is not defined, since
$x^{M-N}$ can tend to either $+\infty$, $-\infty$, or zero, depending on whether
$M>N$, $M<N$, or $M=N$. This implies that Pad\'{e} approximants do not converge for large 
$x \ra \infty$. 

To force a sequence to converge, it is necessary to introduce some quantities that 
regulate the convergence. Such quantities, governing the sequence convergence, can 
be named {\it control functions} The following highlights are the same. The idea of 
introducing control functions so that the reorganized sequence converges was advanced 
in Refs.~\cite{Yukalov_12,Yukalov_13}. Control functions are to be defined by 
{\it optimization conditions} regulating the sequence convergence, so that the initial 
perturbation theory is reorganized to an {\it optimized perturbation theory} yielding 
a converging sequence of optimized approximants~\cite{Yukalov_12,Yukalov_13}. All
mathematical details can be found in recent reviews~\cite{Yukalov_14,Yukalov_15}. 

The introduction of control functions $u_k$ transforms the initial sequence $\{f_k(x)\}$ 
into a sequence $\{F_k(x,u_k)\}$. The latter can be symbolically denoted as
\be
\label{5}
F_k(x,u_k) = \hat R[\; u_k\;] f_k(x) .
\ee
This introduction can be conducted in several ways: by incorporating control functions 
into an initial approximation of perturbation theory, through a change in variables, or 
by a direct sequence transformation~\cite{Yukalov_14,Yukalov_15}. The derivation of equations 
defining control functions starts with the Cauchy criterion of convergence. The criterion  
tells that a sequence converges if and only if, for any $\varepsilon > 0$, there exists a 
number $k_\varepsilon$ such that
\be
\label{6}
|\; F_{k+p}(x,u_{k+p}) - F_k(x,u_k) \;| < \ep ,
\ee
for all $k > k_\varepsilon$ and $p > 0$. 

From the Cauchy criterion, one can derive (see details in~\cite{Yukalov_14,Yukalov_15}) 
the optimization conditions defining the control functions. One possibility is the 
{\it minimal difference} condition 
\be
\label{7}
 F_{k+p}(x,u) - F_k(x,u) = 0 \qquad ( p \geq 1) ,
\ee
whose simplest variant is
\be
\label{8}
  F_{k+1}(x,u) - F_k(x,u) = 0 , \qquad  u = u_k(x)  .
\ee
The other possibility is the {\it minimal derivative} condition   
\be
\label{9}
\frac{\prt F_k(x,u)}{\prt u} = 0 , \qquad  u = u_k(x) .
\ee
Several other representations of optimization conditions are admissible 
\cite{Yukalov_14,Yukalov_15}.

In some cases, control functions can be defined by the boundary conditions
\be
\label{10}
\hat R^{-1}[\; u_k \;] F_k(x,u_k) \simeq f_k(x) \qquad ( x \ra 0 ) ,
\ee
implying asymptotic equivalence at a small variable of the renormalized and 
initial terms of the approximations. The latter condition is also called the 
accuracy-through-order~procedure. 

After the control functions $u_k(x)$ are found from the optimization conditions, the 
optimized approximants read as
\be
\label{11}
f_k^{opt}(x) = \hat R^{-1}[\; u_k\;] F_k(x,u_k(x)) .
\ee

\subsection{Self-Similar Relation}
\label{self}

Self-similar relations are known to arise in renormalization group theory, where one
considers scaling with respect to spatial or momentum variables, as in the real-space 
decimation procedure~\cite{Kadanoff_16,Efrati_17} or in quantum field theory 
\cite{Bogolubov_18,Bogolubov_19}. Then, self-similar relations connect characteristic 
quantities, such as effective Hamiltonians, Lagrangians, or correlation functions, 
with different spatial or momentum scales. In that sense, renormalization groups in 
statistical physics or quantum field theory provide relations between the characteristic 
quantities with scaled variables.  

The notion of self-similarity in approximation theory, introduced in Refs. 
\cite{Yukalov_20,Yukalov_21}, does not concern a scaling of variables, but rather, the 
scaling of approximation orders. The number labeling the approximation order plays the 
role of discrete time. 

It seems natural then to construct a dynamical system in the space of approximations. 
To this end, it is necessary to introduce an endomorphism into the approximation space. 
Let us start with an initial approximation
\be
\label{12}
 f = F_0(x,u_k(x)) , \qquad x = x_k(f) ,
\ee
defining an expansion function $x_k(f)$. Then, it is possible to define the approximation
function
\be
\label{13}
  y_k(f) = F_k(x_k(f),u_k(x_k(f))) .
\ee
The {\it approximation space} is given by the closed linear envelope over all admissible
approximation functions:
\be
\label{14}
 {\cal A} = \overline {\cal L} \{ y_k(f) : ~k = 0,1,2,\ldots \} .
\ee
This is similar to approximation spaces in approximation theory~\cite{Pietsch_22} or 
to phase spaces in physics~\cite{Huang_23,Kubo_24,Bogolubov_25}. Thus, we obtain a dynamical 
system in discrete time, called the {\it approximation cascade}:
\be
\label{15}
 \{ y_k(f) : ~\mathbb{Z}_+ \times {\cal A} \longmapsto {\cal A} \} .
\ee
The points of the cascade form the cascade trajectory.

The Cauchy criterion (\ref{6}) now acquires the form
\be
\label{16}
 |\; y_{k+p}(f) - y_k(f) \; | < \ep .
\ee
If control functions have been chosen so that the sequence of the optimized approximants 
is convergent, then, in terms of dynamical theory, this implies the existence of a fixed
point, where
\be
\label{17}
y_k(y_p^*(f) ) = y_p^*(f).
\ee
Conditions (\ref{16}) and (\ref{17}) lead~\cite{Yukalov_14,Yukalov_15,Yukalov_20,Yukalov_21}
to the {\it self-similar relation}
\be
\label{18}
 y_{k+p}(f) = y_k(y_p(f)) .
\ee
Fixed-point solutions $y_k^*(f)$ to this equation, using the relation (\ref{13}), give 
the self-similar approximants $f_k^*(x) = \hat R^{-1} [u_k] y_k^*(F_0(x,u_k(x)))$. The 
stability of the method can be checked by investigating map multipliers, similarly 
to the stability analysis conducted in numerical calculations~\cite{Richtmyer_26,Golub_27}.  
 
The self-similar relation (\ref{18}) describes the motion of a dynamical system (cascade).
Generally, a dynamical system does not necessarily tend to a fixed point, but it can display
chaotic motion~\cite{Zaslavsky}. This is why the incorporation into the approximation
cascade of control functions, governing sequence convergence, is so important.

\subsection{Self-Similar Root Approximants}
\label{iter}

Generally, the self-similar relation, depending on additional imposed constraints playing 
the role of boundary conditions, can lead to different solutions. For instance, when the 
fixed-point solution is required to satisfy the prescribed asymptotic expansions at small 
as well as at large variables, we obtain~\cite{Gluzman_28} (see also 
\cite{Yukalov_29,Yukalov_30}) the self-similar root approximants 
\be
\label{19}
f_k^*(x) = \left( \left( \left( 1 + A_1 x \right)^{n_1} + A_2 x^2 \right)^{n_2}
+ \ldots + A_k x^k \right)^{n_k}.
\ee
When the large-variable behavior is of the law
\be
\label{20}
 f(x) \simeq B x^\bt \qquad ( x \ra \infty),
\ee
with the known power $\beta$, then we have
\be
\label{21}
 n_j = \frac{j+1}{j} \qquad ( j = 1,2, \ldots,k-1) , 
\qquad 
n_k = \frac{\bt}{k},
\ee
which results in the approximant
\be
\label{22}
 f_k^*(x) = \left( \left( \left( 1 + A_1 x \right)^2 + A_2 x^2 \right)^{3/2}
+ \ldots + A_k x^k \right)^{\bt/k}.
\ee
All control parameters $A_j$ are uniquely defined by the asymptotic boundary condition
\be
\label{23}
 f_k^*(x) \simeq f_k(x)\qquad ( x \ra 0) .
\ee
The root approximant (\ref{22}) at large $x$ values behaves as
\be
\label{24}
 f_k^*(x) \simeq B_k x^\bt \qquad ( x \ra \infty),
\ee
which gives the amplitude
\be
\label{25}
 B_k = \left( \left( \left( A_1^2 + A_2 \right)^{3/2} + A_3\right)^{4/3}
+ \ldots + A_k \right)^{\bt/k}.
\ee
Below, we refer to the root approximants introduced above as iterated root approximants 
or just as iterated roots.

\subsection{Self-Similar Factor Approximants}
\label{fap}

Accepting the asymptotic expansion (\ref{3}) as a boundary condition and representing
this expansion in the form
$$
f_k(x) = \prod_{j=1}^k ( 1 +b_j x)
$$
yields~\cite{Gluzman_3,Yukalov_32} the fixed-point solution in the form of self-similar 
factor approximants
\be
\label{26}
 f_k^*(x) = \prod_{j=1}^{N_k} ( 1 +A_j x)^{n_j},
\ee
where
\begin{eqnarray}
\label{27}
N_k = \left\{ \begin{array}{rl}
k/2 , ~ & ~ k = 2,4, \ldots \\
(k+1)/2 , ~ & ~ k=3,5,\ldots
\end{array} \right.   
.
\end{eqnarray}
For odd $k$ values, scaling relations~\cite{Yukalov_14} allow us to set $A_1 = 1$. 
All other control parameters $A_j$ and $n_j$ are uniquely defined by the boundary 
condition requiring the asymptotic expansion of $f_k^*(x)$ at small $x$ values to 
coincide with $f_k(x)$. The factor approximants serve as a generalized representation of 
the products of functions~\cite{Prudnikov,Gradshteyn}. It has been shown that the 
self-similar factor approximants provide good accuracy for a wide class of problems 
\cite{Gluzman_3,Yukalov_32,Yukalov_5,Yukalova_6}. 

At large $x$ values, the approximant (\ref{26}) gives
\be
\label{28}
 f_k^*(x) \simeq B_k x^{\bt_k} \qquad ( x \ra \infty),
\ee
with the amplitude 
\be
\label{29}
  B_k = \prod_{j=1}^{N_k} A_j^{n_j}  
\ee
and the large-variable exponent
\be
\label{30}
\bt_k = \sum_{j=1}^{N_k} n_j.
\ee

It is appropriate to notice that Pad\'{e} approximants are a particular kind of factor 
approximants, where a portion of the powers equals $+1$ and the other portion is $-1$, 
so that
$$
f_k^*(x) = \frac{\prod_{j=1}^M(1+A_j x)}{\prod_{j=1}^N(1+A_j x)} =
P_{M/N}(x) .
$$

\subsection{Self-Similar Combined Approximants}
\label{comb}

It is possible to combine different types of approximants.  Consider the situation when 
one type of approximation better suits the small-variable limit, while the other type of 
approximation is better in the description of the large-variable limit. For this purpose, 
one can consider the first $q$ terms of the expansion $f_k(x)$ with $q<k$,
\be
\label{31}
 f_q(x) = \sum_{n=0}^q a_n x^n \qquad ( q < k ).
\ee
When constructing a self-similar approximant $f_q^*(x)$ on the basis of expansion 
(\ref{31}), we define the correcting ratio
\be
\label{32}
 C_{k/q}(x) \equiv \frac{f_k(x)}{f_q^*(x)}   
\ee
and expand it in powers of $x$ to obtain
\be
\label{33}
 C_{k/q}(x) \simeq \frac{\sum_{n=0}^k a_nx^n}{\sum_{n=0}^q a_n x^n} \simeq
1 + \sum_{n=q+1}^k a_n x^n.   
\ee
On the basis of the latter expansion, we construct a self-similar approximant 
$C_{k/q}^*(x)$. The final combined approximant is
\be
\label{34}
f_k^*(x) = f_q^*(x) C_{k/q}^*(x).
\ee

As has been mentioned above, Pad\'{e} approximants are a particular kind of self-similar
approximant. So, for the correcting function $C_{k/q}^*(x)$, it is possible to take a
Pad\'{e} approximant $P_{M/N}(x)$ with $M+N=k-q$. In that case, the final approximant is
\be
\label{35}
 f_k^*(x) = f_q^*(x) P_{M/N}(x).
\ee
Of course, the necessary boundary conditions have to be preserved when choosing $q$, $M$, 
\mbox{and $N$.}

\section{Self-Similar Borel-Type Transforms}\label{sec3}

The convergence of series is known to be improved by resorting to Borel 
\mbox{summation~\cite{Borel_2,Borel_33,Glimm_34}.} Borel, or Borel-type summation, 
can be combined by employing self-similar approximants~\cite{YGB}.

\subsection{Self-Similar Borel Transform}

The Borel transform for expansion (\ref{3}) is
\be
\label{36}
 B_k(x) = \sum_{n=0}^k \frac{a_n}{n!} \; x^n,
\ee
with the inverse transformation 
\be
\label{37}
f_k(x) = \int_0^\infty B_k(xt) e^{-t} \; dt .
\ee

The series (\ref{36}) can be summed by employing self-similar approximants, obtaining 
a self-similar Borel transform $B_k^*(x)$. Then, the inverse transformation (\ref{37}) 
gives
\be
\label{38}
f_k^*(x) = \int_0^\infty B_k^*(xt) e^{-t} \; dt.
\ee

The most difficult problem is the study of the large-variable limit. Therefore, we 
pay more attention to this limiting behavior. In the present case, by substituting the 
large-variable form of the self-similar approximant,
\be
\label{39}
 B_k^*(x) \simeq C_k x^{\bt_k} \qquad ( x \ra \infty) ,
\ee
into the integral (\ref{37}), we obtain
\be
\label{40}
f_k^*(x) \simeq B_k x^{\bt_k} \qquad ( x \ra \infty) ,
\ee
where the large-variable amplitude is
\be
\label{41}
 B_k = C_k \Gm(\bt_k + 1).
\ee

\subsection{Self-Similar Borel--Leroy Transform}
\label{bl} 

The Borel-Leroy transform reads as
\be
\label{42}
B_k(x,\lbd) = \sum_{n=0}^k \frac{a_n x^n}{\Gm(n+\lbd+1)} ,
\ee
where $\lambda$ is a control parameter that has to be defined from optimization 
conditions, providing convergence to the sequence of approximants.

The inverse transformation is
\be
\label{43}
 f_k(x) = \int_0^\infty B_k(xt,\lbd)\; t^\lbd e^{-t} \; dt .
\ee
Constructing a self-similar approximation $B_k^*(x,\lambda_k)$ yields the self-similar
optimized Borel--Leroy approximant
\be
\label{44}
f_k^*(x) = \int_0^\infty B_k^*(xt,\lbd)\; t^\lbd e^{-t} \; dt ,
\ee
where the control parameter $\lambda = \lambda_k$ can be defined, e.g., from either the
minimal difference optimization condition (\ref{8}) or the minimal derivative optimization 
condition (\ref{9}) for the large-variable amplitude. 

In the large-variable limit, the form
\be
\label{45}
B_k^*(x,\lbd) \simeq C_k x^{\bt_k} \qquad ( x \ra \infty) 
\ee
results in the approximant (\ref{40}) with the amplitude
\be
\label{46}
 B_k = C_k \Gm(\bt_k + \lbd_k + 1).
\ee

\subsection{Self-Similar Iterated Borel Transform}
\label{ite}

The method used for the Borel and Borel--Leroy transforms can be repeated several times, 
resulting in iterated transforms~\cite{beno}. For example, after accomplishing the first 
step and obtaining the transform
\be
\label{47}
 B_{k1}(x,\lbd^{(1)}) = 
\sum_{n=0}^k \frac{a_n x^n}{\Gm(n+\lbd^{(1)}+1)},
\ee
it is straightforward to repeat the procedure to obtain the double Borel-Leroy transform
\be
\label{48}
 B_{k2}(x,\lbd^{(1)},\lbd^{(2)}) = 
\sum_{n=0}^k \frac{a_n x^n}{\Gm(n+\lbd^{(1)}+1)\Gm(n+\lbd^{(2)}+1)} .
\ee
After $p$ iterations, one obtains
\be
\label{49}
B_{kp}(x,\lbd^{(1)},\lbd^{(2)},\ldots,\lbd^{(p)}) = 
\sum_{n=0}^k \frac{a_n x^n}{\prod_{j=1}^p\Gm(n+\lbd^{(j)}+1)}.
\ee
For simplicity, one assumes that the control parameters at different steps are the same,
so that $\lambda^{(j)} = \lambda$, which gives
\be
\label{50}
B_{kp}(x,\lbd) = 
\sum_{n=0}^k \frac{a_n x^n}{\Gm^p(n+\lbd+1)} .
\ee
The inverse transformation is
\be
\label{51}
 f_k(x) = \int_0^\infty B_{kp}(xt_1t_2t_p,\lbd) 
\prod_{j=1}^p t_j^\lbd \; e^{-t_j} \; dt_j.
\ee

By constructing a self-similar approximant $B_{kp}^*(x,\lambda_{kp})$ on the basis of the 
$p$-iterated transform (\ref{50}), one obtains~\cite{bo} the self-similar iterated
approximants 
\be
\label{52}
  f_{kp}^*(x) = \int_0^\infty B_{kp}^*(xt_1t_2t_p,\lbd_{kp}) 
\prod_{j=1}^p t_j^{\lbd_{kp}} \; e^{-t_j} \; dt_j,
\ee
where the control parameters $\lambda_{kp}$ are defined from the optimization conditions.  
For the large-variable limit of the transform
\be
\label{53}
B_{kp}^*(x,\lbd_{kp}) \simeq C_{kp} x^{\bt_{kp}} \qquad ( x \ra \infty),
\ee
the large-variable behavior of the sought function becomes
\be
\label{54}
 f_{kp}^*(x) \simeq B_{kp} x^{\bt_{kp}} \qquad ( x \ra \infty) ,
\ee
with the amplitude
\be
\label{55}
B_{kp} = C_{kp} \Gm^p(\bt_{kp} + \lbd_{kp} + 1).
\ee

\subsection{Self-Similar Mittag--Leffler Transform}

A generalization of the Borel summation can be conducted by means of the 
Mittag--Leffler~\cite{Mittag_37} summation. By introducing the Mittag-Leffler transform
\be
\label{56}
 M_k(x,\mu) = \sum_{n=0}^k \frac{a_n x^n}{\Gm(n\mu+1)},
\ee
the inverse transformation reads as
\be
\label{57}
 f_k(x) = \int_0^\infty M_k(xt^\mu,\mu) e^{-t} \; dt.
\ee
The series (\ref{56}) can be represented by a self-similar approximant, giving
$M_k^*(x,\mu_k)$, where the control parameter $\mu_k$ is found from the optimization
conditions~\cite{ml}. Then, the inverse transformation yields the self-similar 
approximant
\be
\label{58}  
 f_k^*(x) = \int_0^\infty M_k^*(xt^\mu,\mu) e^{-t} \; dt .
\ee

For large variables, the transform 
\be
\label{59}
M_k^*(x,\mu) \simeq M_k x^{\bt_k} \qquad ( x \ra \infty)
\ee
leads to large-variable behavior of the self-similar approximant
\be
\label{60}
f_k^*(x) \simeq B_k x^{\bt_k} \qquad ( x \ra \infty)   ,
\ee
with the amplitude
\be
\label{61}
 B_k = M_k \Gm(\bt_k \mu_k + 1).
\ee

For the Mittag--Leffler transform, one can define iterated transforms, similarly to the 
iterated Borel--Leroy transforms.

\subsection{Self-Similar Modified Borel--Leroy Transform}

The Borel--Leroy transform can be modified~\cite{Kazakov_39,Kom} to the form
\be
\label{62}
  B_k(x,\lbd,\nu) = a_0 + \sum_{n=1}^k \frac{a_n x^n}{\Gm(n+\lbd+1)n^\nu},
\ee
with the inverse transformation being
\be
\label{63}
  f_k(x) = \int_0^\infty t^\lbd e^{-t}
\left( t \; \frac{\prt}{\prt t} \right)^\nu
B_k(xt,\lbd,\nu) \; dt,
\ee
in which $\nu$ is an integer, so that
\be
\label{64}
 \left( t \; \frac{\prt}{\prt t} \right)^\nu \; x^n = n^\nu x^n.
\ee

Constructing a self-similar transform $B_k^*(x,\lambda_k,\nu)$, where $\lambda_k$ is 
a control parameter, gives~\cite{fr} the inverse transformation
\be
\label{65}
 f_k^*(x) = \int_0^\infty t^{\lbd_k} e^{-t}
\left( t \; \frac{\prt}{\prt t} \right)^\nu
B_k^*(xt,\lbd_k,\nu) \; dt.
\ee
With the large-variable behavior of the transform
\be
\label{66}
B_k^*(xt,\lbd_k,\nu) \simeq C_k x^{\bt_k} \qquad ( x \ra \infty),
\ee
the large-variable limit of the self-similar approximant (\ref{65}) reads as shown in 
Equation (\ref{60}) with the amplitude
\be
\label{67}
  B_k = C_k \bt_k^\nu \; \Gm(\bt_k + \lbd_k + 1).
\ee

As is evident, setting $\lambda_k = 0$ leads to the modified Borel transform.

\subsection{Self-Similar Fractional Iterated Transform}
\label{frd}

The modified Borel--Leroy transform shown in the previous section can be further generalized
in two aspects: First, it is straightforward to iterate it. And, second, it is possible 
to treat the parameters $\nu$ as well as the iteration order $p$, not as integers, but 
as fractional quantities~\cite{bo,fr}. 

Starting with the iterated modified Borel--Leroy transform
\be
\label{68}  
  B_{kp}(x,\lbd,u) = a_0 + \sum_{n=1}^k \frac{a_n x^n}{[\Gm(n+\lbd+1)n^u]^p},
\ee
we have the inverse transformation
\be
\label{69}
  f_{kp}(x) = \int_0^\infty 
\prod_{j=1}^p d t_j \; t_j^\lbd e^{-t_j}
\left( t_j \; \frac{\prt}{\prt t_j} \right)^u
B_{kp}(xt_1t_2\ldots t_p,\lbd,u).
\ee

Then, by constructing a self-similar approximant $B_{kp}^*(x,\lambda_{kp},u_{kp})$, where
$\lambda_{kp}$ and $u_{kp}$ and $p=p_k$ are control parameters, we obtain a
self-similar fractional iterated transform
\be
\label{70}
f_{kp}^*(x) = \int_0^\infty 
\prod_{j=1}^p d t_j \; t_j^{\lbd_{kp}} e^{-t_j}
\left( t_j \; \frac{\prt}{\prt t_j} \right)^{u_{kp}}
B_{kp}^*(xt_1t_2\ldots t_p,\lbd_{kp},u_{kp}).
\ee
Strictly speaking, the differential operator in action (\ref{64}) is defined only for the 
integer $\nu$. However, it is admissible to formally treat $\nu$ as an arbitrary real 
number $u$ when it acts on power law expressions, considering Equation (\ref{64}) as 
a definition.

Then, for the large-variable behavior of the transform
\be
\label{71}
B_{kp}^*(x,\lbd_{kp},u_{kp}) \simeq C_{kp} x^{\bt_{kp}} \qquad 
( x \ra\infty),
\ee
we find the large-variable limit for the sought function
\be
\label{72}
f_{kp}^*(x) \simeq B_{kp} x^{\bt_{kp}} \qquad ( x \ra\infty),
\ee
in which the amplitude is
\be
\label{73}
B_{kp} = C_{kp} \left[\; 
\bt_{kp}^{u_{kp}} \; \Gm(\bt_{kp} + \lbd_{kp} +1 ) \; 
\right]^p.
\ee

In this, and in the previous sections, the control parameters are defined by the 
optimization conditions, as discussed in Section~\ref{cf}, as the minimal difference 
or minimal derivative conditions. In particular, when one is interested in the correct 
evaluation of the large-variable amplitudes, these conditions should be used with respect 
to the amplitudes $B_{kp}(u)$ considered as functions of the control parameters $u$. For 
brevity, here, we use a single control parameter $u$, although there can be several of 
them. It is possible to use the minimal difference condition in the form
\be
\label{74}
B_{k+1,p}(u) - B_{kp}(u) = 0 , \qquad u = u_{kp},
\ee
or in the form
\be
\label{75}
 B_{k,p+1}(u) - B_{kp}(u) = 0 , \qquad u = u_{kp}.
\ee
Otherwise, one can resort to the minimal derivative condition
\be
\label{76}
\frac{\prt B_{kp}(u)}{\prt u} = 0 , \qquad u = u_{kp}.
\ee
For both conditions, the minimal difference and minimal derivative are equivalent. 

Previously, in the paper~\cite{bo}, it was suggested that the number of iterations 
$p$ should be considered a continuous control parameter. The multidimensional integrals 
are relatively easy to define for the integer $p$. Introducing a continuous $p$ implies 
a smooth interpolation between the values of the integrals for discrete $p$ values. The 
approach is applicable only in the limiting case of large $x$ values. 

While setting $u=0$, the parameter $p=p_k$ can be found from the minimal derivative 
condition as the unique solution to the equation
\begin{equation}
\label{77}
\frac{\partial B_{k}(p)}{\partial p}=0.
\end{equation}
Alternatively, we can solve the minimal difference equation 
\be
\label{78}
B_{k+1}(p) - B_{k}(p) =0,
\ee
with a positive integer  $k = 1,2,3,\ldots$.

For the special singular cases of  $\bt=-1,-2\ldots$, one can study the inverse quantity. 
One of the main advantages of the method involving the combination of the methodology shown 
in Section \ref{comb} and the Borel-light summation shown in Section~\ref{light} is that 
they do not involve any explicit singular terms. This property allows one to avoid an extra 
transformation and allows one to work with the original truncations directly. In such cases, 
one is bound to the optimizations with marginal amplitudes, since the complete expression 
for the amplitudes involves singularities.


\section{Critical Amplitudes from Fractional Iterations}
\label{bas}

Consider the case where one has to find explicitly a real, sign-definite, positive-valued 
function $f(x)$ of a real variable $x$. In addition, the function possesses the power law 
asymptotic behavior 
\eqref{4}.
The critical exponent $\bt>0$ is known. The case of a negative $\bt$ 
will be considered by studying the inverse of $f(x)$. Let us find
the amplitude $B$.

We consider the case where it is impossible to find the sought function $f(x)$ explicitly
and exactly from some given equations. 
We show that it is possible to reconstruct the large-variable amplitudes from 
the given truncated asymptotic expansions given in the form \eqref{3}.

Several methods of finding effective sums of the truncated series 
\eqref{1} -\eqref{3} exist, based on the ideas of Borel, Mittag-Leffler, and Hardy 
\cite{beno,sus,Sidi,s1,s2,Soc,bo,ml,fit1,fit2}. The well-known method is that of Pad\'e 
\cite{simp,pa,simp1}. The hypergeometric approximants~\cite{m0,match,me,shal} can be 
employed instead of the Pad\'{e} approximants. More complex are the hypergeometric Meijer 
approximants~\cite{me,shal}. 

Bear in mind that we are interested in accurate analytical calculation of the amplitude 
$B$ in the expression \eqref{4} \cite{ben,fra1,fra2}. To this end, we need, of course, to 
find effective sums for \eqref{1} -\eqref{3}, at least in a general form. The specifics 
of our techniques and the general knowledge of the large-$x$ asymptotic behavior \eqref{4} 
are sufficient in order to find an analytic expression for the amplitudes and to proceed 
by applying the minimal difference and minimal derivative conditions given by Equations 
\eqref{74} -\eqref{78}. 

Let us briefly recapitulate the main tenets of Section~\ref{frd} needed for the 
following applications to concrete problems. First, we set all $\lambda_{kp}$ values to 
zero. This simplification brings us back to a manageable number of control parameters. Below, 
we consider $u$, which is the order of the operator in \eqref{69}, as a continuous control 
parameter. It is possible to define the multidimensional integrals required for calculating  
the critical amplitudes for integers $p$. Introducing continuous $p$ values means a smooth 
interpolation between the values of the integral for discrete $p$ values~\cite{bo,fr}. The 
number of iterations $p$ can also be regarded as a continuous control parameter~\cite{bo}. 
The parameters $\nu$ and $p$ ought to be found from the optimization conditions. The 
general-type optimization conditions are the minimal difference and minimal derivative 
conditions \eqref{74} - \eqref{76}, expressed as the conditions for the critical amplitudes. 
The conditions are equivalent and both are applied below. 

We expect that the solution to the optimization problem exists and is unique. There are 
three main realizations of the Fractional Borel Summation with factor approximants described 
in Section~\ref{fap}. Often, we simply use the term ``Borel summation with factor 
approximants'', but specify the optimization techniques applied in each particular case. 
Below we use the following three methods of finding the critical amplitudes.

\begin{enumerate}
\item 
The first method is based on introducing the fractional order $u$ of the 
differentiation operator. The parameter has to be found from the optimization procedure, 
while $p$ is fixed. The optimization procedure, where only the parameter $u$ is 
considered and $p$ is fixed, is called Fractional Borel Summation with $u$-optimization 
\cite{fr}.
\item  
The second method is based on the optimization of the other parameter $p$, where the
number of iterations extends to arbitrary real numbers from the original integers (see 
\mbox{Section~\ref{ite}} and the paper~\cite{bo}). This has to be found from 
$p$-optimization, while $u$ is fixed. The optimization procedure where only the parameter 
$p$ is considered and $u$ is fixed is called Fractional Borel Summation with $p$-optimization 
\cite{fr}. The optimization is performed according to formulas analogous to 
\eqref{74}--\eqref{76} with straightforward replacement of the parameter $u$ by the 
parameter $p$.
\item  
The third method, Fractional Borel-light or self-similar combined 
approximants is explained in great detail in Sections~\ref{comb} and~\ref{light}. 
The method was suggested in the paper~\cite{fr}, following the main ideas expressed 
earlier in~\cite{YGB,ml}. It is based on optimization with minimal derivative 
or minimal difference conditions of the amplitude $B_{kp}$ (or with the marginal amplitudes 
$C_{kp}$), either with respect to the fractional $u$ or fractional $p$, with subsequent 
correction of the marginal amplitudes with the diagonal Pad\'{e}  approximants  
\cite{YGB,ml,fr,bak}.
\end{enumerate}

We employ some useful formulas for the factor approximants given above in the general 
form in Section~\ref{frd}. These are adjusted to the calculations of the amplitudes 
in Section~\ref{fact}. We also employ the iterated roots explained in Sections
\ref{iter}, and diagonal Pad\'{e} approximants for odd and even number of the terms $k$  
in the truncations~\cite{pa}.

Factor approximants have the advantage of generality, since the case of $\bt=0$ is 
included into the consideration automatically, unlike the case of iterated roots, where 
such a case should be treated individually. On the other hand, iterated roots are very 
user-friendly and can be treated analytically in rather high orders. The problem of 
optimization can be formulated with rather high orders of the perturbation theory, while 
for the factors, only low-order optimizations can be treated analytically.

The diagonal Pad\'{e} approximants are routinely extendable to very high orders. 
Fractional Borel Summation with iterated roots was previously considered in detail~\cite{fr}, 
while the diagonal, odd, and even Pad\'{e} approximants are discussed in~\cite{pa}. 

For completeness and convenience, below, we give some formulas that are required for actual 
computations of the critical amplitudes when the index $\bt$ is known. Two types of 
approximants are discussed in this context, since the iterated roots, presented in  
Section~\ref{iter}, are well adapted to the calculation of the critical 
amplitudes and do not require any specific adjustments.

\subsection{Critical Amplitudes from Factors and Pad\'{e}  Approximants}
\label{fact}

It is both convenient and natural to extrapolate the asymptotic series with power law 
behavior \eqref{4} by means of the self-similar factor approximants 
\cite{Gluzman_3,Yukalov_32}. 
Let us fix the inner sum of the parameters $n_j$ in the Formula \eqref{26} to the known 
index $\bt$, so that
\be
\label{188}
 \bt = \sum_{j=1}^{N_k} n_j
\ee
in all orders. Then, the critical amplitudes can be found by extrapolating the 
series \eqref{1}--\eqref{3} to the form \eqref{26}.
Here
\begin{eqnarray}
\label{133}
N_k = \left\{ \begin{array}{ll}
k/2+1 , ~ & ~ k = 2,4,6,\ldots \\
k/2+1/2 , ~ & ~ k = 1,3,5,\ldots 
\end{array} \right. .
\end{eqnarray}

At large $x$ values, the factor approximants behave as
\be
\label{166}
f^{*}_k(x)\simeq B_k x^{\bt} \qquad ( x \ra \infty ), \;  
\ee
and the amplitude is given by the Formula \eqref{29}. For the even orders $k$, 
one of the $A_j$ values can be set to one. Such a restriction does not change the critical 
index and only influences the value of the amplitude.
From a technical standpoint, it is rather difficult to optimize factor-based methods in 
high orders, although lower orders can easily be optimized.

\subsection{Modified Pad\'{e} Approximations}
\label{pa}

Alternatively, in place of the factors, one can apply the well-known Pad\'{e} approximants 
$P_{n/m}$~\cite{beno}.
The Pad\'{e} approximants can be adapted to calculate the amplitudes at infinity in the 
expression \eqref{4} by means of some transformations. In the case of odd $k=1,3 \ldots$, 
one should study the following transformed series for the sought function $f(x)$, 
$T(x)={f(x)}^{-1/\beta}$.
The following modified Pad\'{e} approximants 
\be
\label{opade}
P_{n,n+1}(x)=\left(PadeApproximant[T[x], {{n, n + 1}}]\right)^{-\beta},
\ee
are defined for even cases, where $n=0,1\ldots,n_{max}$ is a non-negative integer. 

In the odd case, $(2n+1)_{max}=k$.  The approximants \eqref{opade} do comply with the  
power law~\eqref{4} at $x \rightarrow \infty$. 
One can relatively easily find the sequence of approximations 
\begin{equation}
\label{sug1}
B_{n}=\lim_{x\to \infty } \left(x PadeApproximant[T[x], {{n, n + 1}}]\right)^{-\beta},
\end{equation}
for the amplitudes. In the case of even $k=2,4, \ldots$, the following modified-even 
Pad\'{e} approximants 
\be
\label{epade}
P_{n,n}(x)=K(x)\times (PadeApproximant[G[x],n,n]),
\ee 
are defined~\cite{pa}. Here, the corrector  
\be
K(x)=(PadeApproximant[T[x], {{0, 1}}])^{-\beta},
\ee
was introduced to ensure the correct asymptotic behavior~\cite{pa}. Also, 
$
G(x)=\frac{f(x)}{K(x)}
$
represents the transformed truncated series, which can be approached again with the 
diagonal Pad\'{e} approximants. More details on the application of the modified Pad\'{e} 
approximants for the Borel summation can be found in~\cite{pa}. 

The amplitudes sought at infinity can be found as follows:
\begin{equation}
\label{main}
B_{n}=B_{0}\times\lim_{x\to \infty } \left(PadeApproximant[G[x],n,n]\right)
\end{equation} 
with 
$$
B_{0}= \lim_{x\to \infty } \left(K \left(x\right) x^{-\beta} \right).
$$
Here, $n=1,2\ldots, n_{max}$, is a positive integer. In the even case, $n_{max}=k/2$. 
It is not impossible but is very difficult from a technical standpoint to optimize 
Pad\'{e}-based methods, especially at very high orders.

The methods of the factor, root, and Pad\'{e} approximants can be applied individually  
or together in some hybrid form to calculate the marginal amplitude $C$ appearing in 
the course of the Borel transformation. All of the mentioned approximations can be applied 
directly to the series \eqref{1}--\eqref{3} under the asymptotic condition \eqref{1} 
and produce the estimates for the sought amplitude $B$.  Only the iterated roots shown 
in Section~\ref{ite} are seamlessly defined for all $k$ values. The other two approximations 
use two different definitions for the odd and even cases.

\subsection{Critical Amplitudes from Hybrids of Factors and Pad\'{e} Approximations}
\label{light}


The sought function can be reconstructed from the transform $ f_i^*(x)$ 
directly with the help of corrected Pad\'{e} approximants, following the general idea 
of the papers~\cite{ml,YGB}. Such a hybrid approach, coined before Borel-light~\cite{fr}, 
is particularly useful when the integral transformation cannot be applied because 
of singularities in the transformational $\Gm$-functions or when an explicit formula 
and not only numerical values are required. In fact, we are dealing with the whole table 
of approximate values
\be
\label{344}
f_{n,i}^*(x) \simeq  \; P_{n/n}(x)\; f_i^*(x), \,\,\,
\ee
with $i=1,2\ldots k$~\cite{fr}, while $n=1,2\ldots k/2$ for even $k$, and $n=1,2\ldots 
(k-1)/2$ for odd $k$. 
Formula \eqref{34} is built only on the diagonal sequences in adherence with ~\cite{gonch}.

Of course, when only the single approximant $f_q^*(x)$ of the order $q$ with  $q<k$ 
is employed, and only the highest possible order diagonal Pad\'{e} approximant is 
considered, we return to the particular case of the scheme outlined in Section 
\ref{comb}. On the other hand, when $n$ is fixed and $i$ is varied in the Formula 
\eqref{344}, we are dealing with a different sequence of approximations from that described 
in Section~\ref{comb}. The whole (almost) table \eqref{344} was employed in the 
paper~\cite{fr}.

%
%

Since, at $ x \rightarrow \infty$, \,
$f_i^*(x,u)  \simeq C_i(u) x^{\bt},
$
the sought amplitude is approximated by the table of values expressed by means 
of the hybrid formula
\be
\label{marg}
B_{n,i}^*(u_i) = \,P_{n/n}(\infty)\;  C_i(u_i) \,.
\ee
The parameter $u=u_i$ is found from the optimization conditions.
Such conditions are analogous to 
the Equations \eqref{74}--\eqref{76}~\cite{fr} but have to be applied properly to 
the marginal amplitudes  $C_i(u_i)$.


\section{Examples}
\label{U}


In the following sections, we are primarily concerned with the comparison of 
different factor-based approximations with root-based~\cite{fr} and Pad\'{e}-based 
approaches~\cite{pa,and}.

\subsection{Quantum Quartic Oscillator}
\label{OSC}
For the quantum anharmonic oscillator~\cite{Hio1978}, 
perturbation theory yields a rather long expansion for the ground-state energy  
$$
 E_k(g) = \sum_{n=0}^k a_n g^n .
$$
Here, the parameter $g \geq 0$ measures a deviation of the anharmonic potential  from 
the quantum harmonic oscillations.
The coefficients $a_n$ are rapidly growing in magnitude. Their concrete values are known 
for very high orders and can be retrieved from~\cite{Hio1978}. The strong-coupling limit 
for $g \ra \infty$, 
\be
\label{6.7}
 E(g) \simeq B g^{\beta} \qquad
\ee
is also known, and the parameters are $B=0.667986$ and $\beta=1/3$.

Factor approximants, when applied to the truncation $E_k(g)$, give the following sequence 
of approximate amplitudes:
$$ 
B_2=0.678929, \,\, B_3=0.750032,\,\, B_4=0.702102,\,\, B_5=0.724883, 
$$
$$
B_6=0.706184,\,\, B_7=0.712144,\,\, B_8=0.706593,\,\,B_9=0.704931.
$$

Borel factor approximants with $u=0$ and $p=1$
give 
$$ 
B_2^*=0.708098, \,\, B_3^*=0.742158,\,\, B_4^*=0.689432,\,\, B_5^*=0.688555,
$$
$$
B_7^*=0.672477,\,\,B_8^*=0.679915,\,\,B_9^*=0.677194.
$$
Complex values are omitted here and in what follows.

The results produced by the following Borel-light approximations with a fixed marginal 
amplitude and varying correcting terms,
$$
B_{2,7}^*(1) = \; P_{2/2}(\infty)\;  C_7(1)=0.626631,
B_{3,7}^*(1) = \; P_{3/3}(\infty)\;  C_7(1)=0.687644,
$$
$$
B_{4,7}^*(1) = \; P_{4/4}(\infty)\;  C_7(1)=0.690745,\,\,\,\,
B_{5,7}^*(1) = \; P_{5/5}(\infty)\;  C_7(1)=0.690887,
$$
are slightly worse.

The best results are found for rather high orders with the following Borel-light 
approximations with varying marginal amplitudes and correcting terms of the same order,
$$
B_{13,2}^*(1) = \; P_{13/13}(\infty)\;  C_2(1)=0.69341,
$$
$$
B_{13,4}^*(1) = \; P_{13/13}(\infty)\;  C_4(1)=0.68615,
$$
$$
B_{13,5}^*(1) = \; P_{13/13}(\infty)\;  C_5(1)=0.68766,
$$
$$
B_{13,7}^*(1) = \; P_{13/13}(\infty)\;  C_7(1)=0.67907,
$$
$$
B_{13,8}^*(1) = \; P_{13/13}(\infty)\;  C_8(1)=0.68904,
$$
$$
B_{13,9}^*(1) = \; P_{13/13}(\infty)\;  C_9(1)=0.67397.
$$

Bear in mind that, for the factor approximants, the optimization problems can be solved 
analytically only for low orders. Only the minimal derivative problem possesses a unique 
solution. The results of optimization, such as the optimal control parameters, can be used 
as inputs to construct the sequence of Borel-light approximations. But, even in these cases, 
the results appear not to be better than those given by the factor approximants by 
themselves. However, such a strategy can be useful for some other problems and is 
exploited~below.

Various other methods are considered as well. They give even more reasonable results, 
as shown in Table~\ref{Table 1}.
\begin{table}[hp]
\caption{Critical amplitude for the one-dimensional quartic oscillator}
\label{Table 1}   
\vskip 2mm   
\centering
\begin{tabular}{cc} \toprule
\textbf{Quartic Oscillator} & \textbf{Amplitude} \\  \midrule
Factor approximants         &  0.704931  \\ \midrule
Borel with factors          &  0.677194  \\ \midrule
Borel-light with factors    & 0.690887  \\ \midrule
Exact                       & 0.667986 \\ \midrule
Fractional Borel with roots~\cite{fr}  & \bf{0.670902} \\ \midrule
Borel with roots~\cite{fr}             & 0.682494 \\   \midrule
Even Pad\'{e}--Borel~\cite{pa}         & 0.679037\\  \midrule
Odd Pad\'{e}--Borel~\cite{pa}          & 0.67926\\ \midrule
Even Pad\'{e}~\cite{pa}                & 0.709572\\ \midrule
Odd Pad\'{e}~\cite{pa}                 & 0.712286\\ \bottomrule
\end{tabular}
\end{table}
The best result is marked in bold in Table~\ref{Table 1}.
 
It is achieved  in the ninth 
order of perturbation theory  with the Fractional Borel Summation with iterated 
roots~\cite{fr}. An even better result, $B=0.669356$, is found with the same method in 
the $10$th order~\cite{fr}.

The method of Corrected Pad\'{e} approximants needs more terms to obtain the same accuracy 
\cite{cor} but is able to easily scan very high orders~\cite{cor}. But, already in 
the $10$th order, it gives a reasonable estimate,
$B=0.655086$.

\subsection{Schwinger Model: Energy Gap}
\label{GAP}
Let us consider the energy gap between the lowest and second excited states of the scalar 
boson for the massive Schwinger model in Hamiltonian lattice theory~\cite{Schwinger_15,Hamer_16}. 

The energy gap  $\Delta(z)$ between the two states at small $z$ values can be represented 
(in low orders), according to the paper~\cite{Hamer_16}, as
\be
2 \Dlt(z) \simeq
1+6 z-26 z^2+190.6666666667 z^3-1756.666666667 z^4+18048.33650794 z^5,
\ee
with the variable $z = (1/ga)^4$. Here, $g$ stands for a coupling 
parameter and $a$ is the lattice spacing. The coefficients $a_n$ are known up to 
the $13$th order and can be found in~\cite{Hamer_16}.

In the continuous limit, 
$\Dlt(z)$ follows the power law~\cite{Hamer_16}
\be
\label{04}
\Dlt(z) \simeq  B z^{\bt} \qquad ( z \ra \infty),
\ee
where $B=1.1284$ and $\bt=1/4$.

Factor approximants, when applied to the original expansion, give the following sequence:
$$ B_2=1.58276, \,\, B_3=1.574592,\,\, B_5=1.45957,\,\, B_7=1.36643,$$ 
$$ B_8=1.05181,\,\,B_9=1.3421,
$$
with a reasonable value given by $B_8$. The complex values are omitted again here and 
below.

Borel factor approximants, calculated with $u=0$ and $p=1$, give the following sequence:
$$ B_2^*=2.02627, \,\, B_3^*=1.59039,\,\, B_4^*=1.55169,\,\, B_5^*=1.58126,$$
$$
B_6^*=1.62346,\,B_7^*=1.59431,\,\,B_8^*=1.05511,
$$
with a reasonable result that is very similar to the previous result $B_8$.

The most consistent results are found by applying the Borel-light approximations 
with a fixed marginal amplitude and varying correcting terms,

$$
B_{4,2}^*(1) = \; P_{4/4}(\infty)\;  C_2(1)=1.32253,
$$
$$
B_{5,2}^*(1) = \; P_{5/5}(\infty)\;  C_2(1)=1.05724,
$$
$$
B_{6,2}^*(1) = \; P_{6/6}(\infty)\;  C_2(1)=1.15488.
$$
The last two terms can be considered lower and upper bounds, respectively.

As mentioned previously, the optimization problem with factor approximants can be solved 
analytically for low orders only. The optimization results, such as those for the optimal 
control parameters, can be used to construct the sequence of Borel-light approximations. 
Such a strategy appears to be useful, as shown below. Indeed, the application of fractional 
Borel $u$-optimization with factor approximants for $p=1,2$  amounts to solving the equation
$$
B_{3,2}\left(u\right) - B_{3,1}\left(u\right)=0.
$$
It brings a sensible result for the sought amplitude $B=1.21118$ with the control 
parameter $u=u_3=0.286426$.

The Borel-light approximations with the same marginal amplitude and varying correcting 
terms give even better results:
$$
B_{4,3}^*(u_3) = \; P_{4/4}(\infty)\;  C_3(u_3)=1.18766,
$$
$$
B_{5,3}^*(u_3) = \; P_{5/5}(\infty)\;  C_3(u_3)=1.22814,
$$
$$
B_{6,3}^*(u_3) = \; P_{6/6}(\infty)\;  C_3(u_3)=1.18951.
$$

Different resummation methods give the results presented in Table~\ref{Table 2}.
The method of Fractional Borel-light summation with 
iterated roots~\cite{fr} also gives a rather good result: $B=B_{6,10}=1.1452$. It is rather 
close to the value $B=1.14(3)$ obtained with finite-lattice calculations. Bear in mind that 
different advanced series methods give $B=1.25(15)$~\cite{Hamer_16}.

\begin{table}[hp]
\caption{Schwinger model-gap}
\label{Table 2} 
\vskip 2mm      
\centering
\begin{tabular}{cc}
\toprule
\textbf{Schwinger Model} &  \textbf{Gap}   \\ \midrule
Factor approximants, $8$th order        & 1.0518 \\ \midrule
Factor approximants, $9$th order        & 1.3421 \\ \midrule
Borel with factors                      & 1.0551  \\ \midrule
Borel-light with factors                & 1.1549  \\ \midrule
Borel-light with factors, $u$-optimal   & 1.1895\\ \midrule
Exact                                   & 1.1284   \\ \midrule
Borel with roots (average)~\cite{bo}    & \bf{1.1224} \\  \midrule
Odd Pad\'{e}, $11$th order~\cite{pa}    & 1.2266  \\ \midrule
Corrected Pad\'{e} approximants~\cite{cor}  & 1.2468\\ \midrule
Borel-light with roots ~\cite{fr}           & 1.1452\\ \bottomrule
\end{tabular}
\end{table}
The best result is marked in bold in Table~\ref{Table 2}.

\subsection{Schwinger Model: Critical Amplitude}
\label{Sch2}


The ground-state energy $E$ of the Schwinger model is given by the very short truncated 
series \cite{Carrol_29,Vary_30,Adam_31,Striganesh_32,Coleman_33,Hamer_34,Hamer_16},
\begin{equation}
\label{377}
 E(x) \simeq 0.5642 - 0.219 x + 0.1907 x^2 \qquad ( x \rightarrow 0 ) .
\end{equation}
The large-$x$ behavior is also known. It is expressed in the form shown in \eqref{4},
$$E(x) \simeq B x^{\bt} \qquad ( x \rightarrow \infty ),$$
with $B=0.6418$, $\bt=-1/3$. Sometimes, an addition of one more trial term with $a_3=0$ 
may help to improve the results. 

Because of the negative $\bt$, we work with the inverse truncations when using 
$u$-optimization. The problem appears to be quite difficult for the factor 
approximations, i.e., the factor approximant of the second order gives only $B\approx 0.5456$.  
The application of fractional Borel $u$-optimization with the factor approximants for 
$k=3$, $p=1, 2$ amounts to solving the equation
$$
B_{3,2}\left(u\right) - B_{3,1}\left(u\right)=0.
$$
It brings, after the inversion, a very good result for the sought amplitude $$B=0.642257$$ 
with the uniquely determined control parameter $u=u_3=0.0961685$. The result is the best 
among all of the results represented in Table~\ref{Table 3}. It is even better than the 
result found from the self-similar root approximants, which explicitly employs the known 
subcritical index~\cite{YGh}.  

A $u$-optimal Borel-light technique with factors applied to the inverse quantities
gives the inverse amplitude 
$$
B_{2,3}^*(u_k) = \; P_{2,2}(\infty)\;  C_3(u_3), \,\,\,
$$
and the total critical amplitude $B$ after inversion equals $0.5827$. 

For $p$-optimization with $u=0$, the parameter $p=p_3$ is found from the minimal 
derivative condition as the unique solution to the equation
$$
\frac{\partial B_{3}(p)}{\partial p}=0,
$$
with $p=p_3=0.897973$ and $B=B_3=0.637067$.

As expected, the results found with optimizations and presented above represent an improvement 
over the non-optimized Borel factor techniques corresponding to $u=0$ and $p=1$, which also 
produce the very reasonable $B\approx 0.6351$.

The results of calculations by various methods are shown in Table~\ref{Table 3}. 
\begin{table}[hp]
\caption{Schwinger model: amplitude}
\label{Table 3}  
\vskip 2mm     
\centering
\begin{tabular}{cc}
\toprule
\textbf{Schwinger} & \textbf{Amplitude}   \\ \midrule
Factor approximant, second order       & 0.5456  \\ \midrule
Factor approximant, third order        & complex  \\ \midrule
Fractional Borel with factors, ($p=1,2$), $u$-optimal  & \bf{0.6423}  \\ \midrule
Borel with factors ($u=0$, $p=1$)                      & 0.6351  \\ \midrule
Fractional Borel with factors, $p$-optimal, min.derivative   & 0.6371  \\ \midrule
Fractional Borel with factors, $p$-optimal, min.difference   & 0.5639  \\ \midrule
Borel-light with factors, $u$-optimal                        & 0.5827  \\ \midrule
Exact                                                        & 0.6418   \\ \midrule
Fractional Borel with roots ~\cite{fr}                       & 0.6672  \\ \midrule
Odd Pad\'{e}--Borel~\cite{pa}                                & 0.6122  \\ \midrule
Borel with roots ~\cite{fr}             & 0.6562  \\ \midrule
Iterated root, second order & 0.5523 \\ \midrule
Odd Pad\'{e}~\cite{pa}       & 0.5344   \\ \bottomrule
\end{tabular}
\end{table}
The best result is marked in bold in Table~\ref{Table 3}.

It is worth stressing that it is always useful to attack the problem using several methods. 

\subsection{Anomalous Dimension}
\label{ANOM}

Consider the cusp anomalous dimension $\Omega(g)$ of a light-like Wilson loop in the $n=4$ 
supersymmetric Yang--Mills theory ~\cite{ban,YGh}.
It depends only on the variable $x=g^2$ expressed though
the coupling $g$. The problem can be written down in terms of the function  
$
f(x)=\frac{\Omega(x)}{x}.
$
The latter function has the following weak-coupling expansion,
$$ f(x) \simeq 4-13.1595 x+95.2444 x^2-937.431 x^3,  \quad x \rightarrow 0.$$
In the strong-coupling limit, $f(x)$ takes the form shown in \eqref{4}, 
$$f(x) \simeq B x^{\bt}, \quad x \rightarrow \infty ,  $$
with $B=2$ and $\beta=-1/2$.

Direct application of the factor approximants in the second order gives $B \approx 2.1307$, 
while in the third order, it gives $B\approx 1.8389$. The Borel summation with factor 
approximants in the second order gives $B\approx1.8798$ and is better than the direct factor 
approximation. In the third order, the Borel summation with factors gives complex 
results. 

Neither of the optimization types brings a unique solution. The best result, 
$B\approx 2.0233$, is found with $p$-optimization with $u=0$,
$$
B_{3}\left(p\right) - B_{2}\left(p\right)=0,
$$
with the optimum obtained at $p=p_2 \approx 0.43297$.
  
The Borel-light technique, when applied to inverse quantities, gives the inverse amplitude 
$$
B_{2,2}^*(p_2) = \; P_{2/2}(\infty)\;  C_2(p_2), \,\,\,
$$
and the total critical amplitude after inversion is $B=1.92018$. 
meaning is retained.

The result $B\approx 2.1115$ is found for the $u$-optimization
$$
B_{3,1}\left(u\right) - B_{2,1}\left(u\right)=0 ,
$$
with the optimal value $u=u_2 \approx 0.263034$.

The fractional Borel technique with iterated roots~\cite{fr} gives a unique solution 
$B=1.90291$ in the case of the $u$-optimization problem. 
  
Some results obtained by different methods are shown in Table~\ref{Table 4}. The best 
result, $B=2.0118$, is obtained from the methodology described above in Section~\ref{bl}. 

\begin{table}[hp]
\caption{Cusp}
\label{Table 4}      
\vskip2mm
\centering
\begin{tabular}{cc}
\toprule
\textbf{Cusp} & \textbf{Amplitude}  \\ \midrule
Factor approximant, second order          & 2.1307\\ \midrule
Factor approximant, third order           & 1.8389\\ \midrule
Borel with factors, second order          & 1.8798 \\ \midrule
Borel with factors, third order           & complex \\ \midrule
Borel-light with factors, $p$-optimal     & 1.9202\\ \midrule
Exact                                     & 2\\ \midrule
Fractional Borel with roots~\cite{fr}     & 1.9029 \\ \midrule
Odd Pad\'{e}~\cite{pa}                    & 1.7973  \\ \midrule
Optimal Borel--Leroy~\cite{ml}            & {\bf{{2.0118}}} \\ \midrule 
Borel  with roots   ~\cite{fr}            & 2.4416 \\ \midrule
Iterated Roots                            & 1.6977\\ \bottomrule 
\end{tabular}
\end{table}
The best result is marked in bold in Table~\ref{Table 4}.

\subsection{Two-Dimensional Polymer: Swelling}
\label{2dp}

For a two-dimensional polymer one can theoretically study the so-called swelling factor 
$\Upsilon$ \cite{Muthukumar_1}. For $\Upsilon$, perturbation theory yields expansions in 
the powers of the dimensionless coupling parameter $g$. 
The swelling 
factor can be represented as the following truncation,
\begin{equation}
\label{6333}
\Upsilon(g) \simeq 1 + \frac{1}{2} \; g - 0.12154525 \; g^2+ 0.02663136 \,
g^3-0.13223603\,g^4\, \; (g\rightarrow 0).
\end{equation}
As $g\rightarrow \infty$, the swelling factor is expressed in the form of \eqref{4}, i.e., 
$$
\Upsilon(g)\simeq B g^\bt .
$$ 
Here, the critical index $\bt=1/2$ is exact~\cite{grosb,pel}. 
As for the amplitude $B$, one can only say that it is of the order of unity.

Indeed, factor approximants and Borel factor techniques all give results close to unity. 
For $p$-optimization with $u=0$, the parameter $p$ can be found from the minimal derivative 
condition
$$
\frac{\partial B_{3}(p)}{\partial p}=0 ,
$$
with $p=p_3=0.414668$ and $B=B_3=0.982576$. A much lower result, $B\approx 0.82498$, is 
found with $u$-optimization, 
$$
B_{3,2}\left(u\right) - B_{3,1}\left(u\right)=0,
$$
with the unique optimum located at $u=u_3 \approx 0.443811$. The Borel-light summation 
technique fails to produce a holomorphic diagonal Pad\'{e} approximation.

The results are presented in Table~\ref{Table 5}. Most of the methods give results that 
are close to the conjectured value of unity. Even Pad\'{e} approximants, factor approximants, 
and Borel summation with factors all give values close to one. 

\begin{table}[hp]
\caption{Two-dimensional polymer}
\label{Table 5}     
\vskip 2mm
\centering
\begin{tabular}{cc}
\toprule
$2D$ \textbf{Polymer}    & \textbf{Critical Amplitude}   \\ \midrule
Factor approximant, third order             & 1.0004\\ \midrule
Factor approximant, fourth order            & 1.00006\\ \midrule
Borel with factors, third order,  ($p=1$, $u=0$)  & 1.00734  \\ \midrule
Borel with factors, fourth order,  ($p=1$, $u=0$) & 0.97209  \\ \midrule
Fractional Borel with factors, min. derivative, $p$-optimal & 0.98258 \\ \midrule
``Exact'' conjectured                                       & 1\\ \midrule 
Fractional Borel  with roots~\cite{fr}                      & 0.970678 \\ \midrule
Even  Pad\'{e}~\cite{pa}                                    & \bf{1.00002}  \\ \midrule
Even  Pad\'{e}--Borel~\cite{pa}                            & 0.977767  \\ \midrule
Borel  with roots ~\cite{fr}                               & 0.9696  \\ \midrule
Iterated Roots    & 0.970718  \\ \bottomrule   
\end{tabular}
\end{table}
The best result is marked in bold in Table~\ref{Table 5}.

\subsection{Three-Dimensional Polymer: Swelling}
\label{3dp}
For a three-dimensional polymer, one can find the swelling factor $\Upsilon(g)$
\cite{Muthukumar_1,Muthukumar_2}, in the form of
a truncated series of the type \eqref{3}, 
\begin{equation}
\label{633}
\Upsilon(g) \simeq 1 + \frac{4}{3} \; g -2.075385396 \; g^2+6.296879676\,g^3-25.05725072\,g^4\, \;(g\rightarrow 0).
\end{equation}
The expansion \eqref{633} can be extended to the sixth order~\cite{Muthukumar_1,Muthukumar_2}.
The strong-coupling behavior of the expansion factor is expressed in the form of \eqref{4},
$$
\Upsilon(g)\simeq B g^\bt \; (g\ra \infty).
$$
with $B\approx 1.531$, 
and $\bt\approx 0.3544$. 

Factor approximants give the following sequence:
$$ B_2=1.50647, \,\, B_3=1.54784,\,\, B_4=1.53523,\,\, B_5=1.53983, \,\,B_6=1.53701 .$$
Borel factor approximants give the sequence
$$ B_2^*=1.60365, \,\, B_3^*=1.55123,\,\, B_4^*=1.51916,\,\, B_5^*=1.53117,\,\,B_6^*=1.53992 \;.$$

For $p$-optimization with $u=0$, the parameter $p=p_3$ can be found from the 
condition
$\frac{\partial B_{3}(p)}{\partial p}=0.
$
The unique solution to the latter equation,
$p=p_3=0.358042$, is found and the critical amplitude is $B=B_3=1.53441$.  

The Borel-light technique gives  
$$
B_{2,3}^*(p_3) = \; P_{2/2}(\infty)\;  C_3(p_3)=1.53998,
$$
$$
B_{3,3}^*(p_3) = \; P_{3/3}(\infty)\;  C_3(p_3)=1.54002.
$$
A much lower result, $B=1.17626$, is found 
with $u$-optimization, 
$$
B_{3,1}\left(u \right) - B_{2,1}\left(u \right),
$$
with the optimal value $u=u_2=0.369247$. However, the Borel-light technique gives  
$$
B_{2,3}^*(u_2) = \; P_{2/2}(\infty)\;  C_3(u_2)=1.47006,
$$
$$B_{3,3}^*(u_2) = \; P_{3/3}(\infty)\;  C_3(u_2)=1.53475,
$$
and the latter estimate for the critical amplitude is rather good.

The results are shown in Table~\ref{Table 6}. 

\begin{table}[hp]
\caption{Three-dimensional polymer}
\label{Table 6}     
\vskip2mm
\centering
\begin{tabular}{cc}
\toprule
$3D$ \textbf{Polymer} & \textbf{Critical Amplitude}   \\ \midrule
Factor approximant, fifth order  &1.5398\\ \midrule
Factor approximant, sixth order  &1.537\\ \midrule
Borel factors, fifth order  &{\bf{{1.5312}}}\\ \midrule
Borel factors, sixth order  &1.5399\\ \midrule
Borel-light with factors, sixth order, $p$-optimal & 1.54\\ \midrule
Borel-light with factors, sixth order, $u$-optimal & 1.5348\\ \midrule
``Exact '' numerical ~\cite{Muthukumar_2}             &1.5309\\ \midrule
Fractional Borel with roots~\cite{fr}& 1.53523 \\ \midrule
Even  Pad\'{e}, sixth order~\cite{pa}       & 1.54022  \\ \midrule
Even  Pad\'{e}--Borel~\cite{pa}       & 1.53296  \\ \midrule
Odd Pad\'{e}, fifth order ~\cite{pa}       & 1.54089  \\ \midrule
Odd  Pad\'{e}--Borel~\cite{pa}       & 1.52996  \\ \midrule
Iterated Roots,  sixth order              & 1.53611  \\ \midrule
Borel with roots    & 1.52718  \\ \bottomrule
\end{tabular}
\end{table}
The best result is marked in bold in Table~\ref{Table 6}.

The fractional $p$-optimal Borel technique with roots~\cite{fr} gives a fairly reasonable 
value, $B=1.53523$. 
Consistent results have also been found with the Pad\'{e}--Borel techniques 
shown in the paper~\cite{pa}. 

\subsection{One-Dimensional Quantum Nonlinear Model}
\label{nonl}
The ground-state energy of 
the Bose-condensed atoms in a harmonic trap can be expressed in 
terms of the function $f(g)$, which can be expanded 
with a dimensionless coupling parameter $g$. For the fifth order,
$
f_5(g) = 1 + \sum_{n=1}^5 a_n z^n,
$
where all $a_n$ values can be found in the paper~\cite{Yuk2012}. The coefficients rapidly 
decay by the absolute value. 
The strong-coupling limit is given in the form \eqref{4},
$
f(g) \simeq \frac{3}{2}\, g^{2/3}.
$

Factor approximants give the following, apparently convergent, sequence for the amplitude:
$$ B_2=1.46572, \,\, B_3=1.49615,\,\, B_4=1.49306,\,\, B_5=1.49188.$$

Borel factor approximants give the sequence
$$ B_2^*=1.52317, \,\, B_3^*=1.6084,\,\, B_4^*=1.59098,\,\, B_5^*=1.42684,$$
which defines the upper and lower bounds.

For $p$-optimization with $u=0$, the parameter $p=p_3$ can be found as the unique solution 
to the equation
$$
\frac{\partial B_{3}(p)}{\partial p}=0,
$$
with $p=p_3=-0.605018$ and $B=B_3=1.47151$. The Borel-light technique gives  
$$
B_{2,3}^*(p_3) = \; P_{2/2}(\infty)\;  C_3(p_3), \,\,\,
$$
leading to fairly reasonable numbers, $B=1.49291$, for the the critical amplitude.

A lower value, $B\approx 1.42983$, is found 
with $u$-optimization from the equation
$$
B_{3,2}\left(u\right) - B_{3,1}\left(u\right)=0,
$$
with the optimal value $u=u_3 \approx 0.653582$. The Borel-light technique gives  
$$
B_{2,3}^*(u_3) = \; P_{2/2}(\infty)\;  C_3(u_3), \,\,\,
$$
and the critical amplitude, $$B=B_{2,3}^*(u_3) = 1.50763,$$ is very good.

The results are presented in Table~\ref{Table 7}. Almost all of them have a good level 
of accuracy.

\begin{table}[hp]
\caption{Non-linear quantum model}
\label{Table 7}     
\vskip2mm
\centering
\begin{tabular}{cc}
\toprule
$1d$ \textbf{Non-Linear Model} & \textbf{Amplitude}   \\ \midrule
Factor approximant, $4$th order  &1.4931\\ \midrule
Factor approximant, $5$th order  &1.4919\\ \midrule
Borel factors, $4$th order  &1.591\\ \midrule
Borel factors, $5$th order  &1.4268\\ \midrule
Borel-light with factors, $4$th order, $p$-optimal & 1.4922\\ \midrule
Borel-light with factors, $4$th order, $u$-optimal & {\bf{{1.5076}}}\\ \midrule
Exact               & 3/2   \\ \midrule
Fractional Borel with roots~\cite{fr}          & 1.4759 \\ \midrule
Odd  Pad\'{e}~\cite{pa}      & 1.4923  \\ \midrule
Even  Pad\'{e} ~\cite{pa}       & 1.4918 \\ \midrule
Borel with roots   ~\cite{fr}         & 1.3851  \\ \midrule
Iterated roots           & 1.448  \\ \bottomrule
\end{tabular}
\end{table}
The best result is marked in bold in Table~\ref{Table 7}.
%

Particularly good results are obtained with the Fractional Borel-light optimal techniques. 
The Pad\'{e} techniques described in the paper~\cite{pa} also give good results, improving 
the techniques of iterated roots.

\subsection{Three-Dimensional Harmonic Trap}
\label{Trap}

The ground-state energy $E$ of the trapped Bose condensate in the three-dimensional case 
was investigated in the paper~\cite{BCY}. The energy  can be approximated by the following
truncation
\begin{equation}
\label{spw}
E(c)\simeq \frac{3}{2}+\frac{1}{2} c -\frac{3}{16} c^2+\frac{9}{64} c^3 -\frac{35}{256} c^4 
\qquad(c\rightarrow 0) ,
\end{equation}
with the ``trapping'' parameter $c$.

For a very strong parameter $c$, the energy behaves as the power law
\begin{equation}
\label{sps}
E(c)\simeq B c ^{\bt} 
\qquad (c\rightarrow \infty,
\end{equation}
And, the critical parameters $B=\frac{5}{4}$, $\bt=2/5$~\cite{BCY} are known.

Factor approximation in the second order fails, while in the higher-order factor, the
approximants give reasonably good results. However, the Borel summation with the factor 
approximants in the second order works very well, with $B \approx 1.25983$.

The use of fractional Borel $u$-optimization with the factor approximants 
for $k=3$, $p=1, 2$ amounts to solving the equation
$$
B_{3,2}\left(u\right) - B_{3,1}\left(u\right)=0.
$$
It gives an amplitude of $B=1.15323$ with the control parameter $u=u_3=0.160358$. 
The corrected Borel-light technique gives the amplitude 
$$
B_{2,3}^*(u_3) = \; P_{2/2}(\infty)\;  C_3(u_3), \,\,\,
$$
and the total critical amplitude $$B=1.2561,$$ is the best among all estimates.

The result, $B\approx 1.2911$, is found with $p$-optimization with $u=0$, which 
amounts to solving the equation
$$
B_{3}\left(p\right) - B_{2}\left(p\right)=0.
$$
The optimum is found for $p=p_2=0.705782$. The Borel-light technique gives the amplitude 
$$
B_{2,2}^*(p_2) = \; P_{2/2}(\infty)\;  C_2(p_2), \,\,\,
$$
and the total critical amplitude, $B=1.28598$, is quite reasonable.
The results from the calculations using different methods are presented in Table~\ref{Table 8}. 
All of them give close and rather accurate values.

\begin{table}[hp]
\caption{Trap}
\label{Table 8}   
\vskip 2mm  
\centering
\begin{tabular}{cc}
\toprule
\textbf{$3d$ Trap} & \textbf{Amplitude}   \\ \midrule
Factor approximant, third order              &1.30227\\ \midrule
Factor approximant, fourth order              &1.2848\\ \midrule
Borel with factors, second order              &1.2598\\ \midrule
Borel with factors, third order            & 1.2916  \\ \midrule
Borel with factors,  fourth order            & 1.2858  \\ \midrule
Fractional Borel with factors, min.diff. $p$-optimal           & 1.2911\\ \midrule
Borel-light  with factors, $u$-optimal            & {\bf{{1.2561}}}  \\ \midrule
Borel-light  with factors, $p$-optimal            & 1.286  \\ \midrule
Exact               &5/4\\ \midrule
Fractional Borel with roots ~\cite{fr}       & 1.2852\\ \midrule
Even  Pad\'{e}~\cite{pa}       & 1.28211 \\ \midrule
Even  Pad\'{e}--Borel~\cite{pa}       & 1.2855  \\ \midrule
Borel with roots  ~\cite{fr}          & 1.28492  \\ \midrule
Iterated roots             & 1.2739  \\ \bottomrule
\end{tabular}
\end{table}
The best result is marked in bold in Table~\ref{Table 8}.

In both cases of trapped Bose condensates discussed above, the methods based on the 
idea of corrected Borel-factor-light approximants work well. They give numbers that are 
better than those of the Pad\'{e}  techniques or those of methodologies based on iterated 
roots.

\subsection{Bose Temperature Shift}
\label{BS}

The ideal Bose gas is unstable below the condensation temperature $T_0$~\cite{BCY}. Atomic 
interactions induce the shift $\Dlt T_c \equiv T_c - T_0$ to the realistic $T_c$ of a 
non-ideal Bose system. The shift is characterized by the ratio 
$$
\frac{\Dlt T_c}{T_0} \simeq c_1 \gm \; ,
$$
for the asymptotically small gas parameter $\gm \ra 0 $.


Monte Carlo simulations~\cite{Arn2001a, Arn2001b, Nho2004} suggest that
\be
\label{7.9}
c_1 = 1.3. \pm 0.05.
\ee
In order to calculate $c_1$ theoretically, it has been suggested that one should first 
calculate an auxiliary function $c_1\left(g\right)$~\cite{Kas2004a, Kas2004b, Kas2004c}. 
Then, one can find $c_1$ as follows:
\be
\label{7.10}
c_1 = \lim_{g\ra\infty} c_1\left(g\right) \equiv B .
\ee
The latter limit is found from the expansion 
over an effective coupling parameter,
\be
\label{7.11}
c_1\left(g\right) \simeq =0.223286 g + -0.0661032 g^2 + 0.026446 g^3 -0.0129177 g^4 + 0.00729073 g^5.
\ee

In the fourth order of factor approximants, we find $c_1\approx 1.1$, which is much smaller 
than that expected from the simulations. 

In what follows, we work with the original $c_1(g)$.
For instance, in $p$-optimization with $u=0$, we can set, by analogy with 
Equation \eqref{78},
\be
C_{3}\left(p\right) - C_{2}\left(p\right)=0.
\ee
The latter equation gives the optimal solution $p=p_2 = 0.41657$. Using the Borel-light 
technique with the optimal $p_2$, we have a good result
$$B=B_{2,2}^*(p_2) = P_{2/2}(\infty)\;  C_2(p_2)=1.28421 .$$
Even by just setting $p=1$, $u=0$, we arrive at
$$B=B_{2,2}^*(1) =  \; P_{2/2}(\infty)\;  C_2(1)= 1.17351 \;.$$
The application of fractional Borel $u$-optimization with factor approximants  
amounts to solving the equation
\be
\label{another}
C_{3,1}\left(u\right) - C_{2,1}\left(u\right)=0 ,
\ee
which is written in analogy with \eqref{75}. It gives the control parameter 
$u=u_2=-0.4351064$. The correction with the Borel-light technique gives the amplitude 
$$
B_{2,2}^*(u_2) = \; P_{2/2}(\infty) \,  C_{2}\left(u_2\right), 
$$
where $C_{2}\left(u_2\right)\equiv C_{2,1}\left(u_2\right)$, and the sought amplitude is 
$B=B_{2,2}^*=1.27224$. 

Note that there is another solution to the equation \eqref{another}, $u_2=-0.56557$, 
and it gives $B_{2,2}^*=1.28553$. The latter result appears to be identical to the results 
of Modified Even Pad\'{e} summation~\cite{pa}. The two solutions are very similar to each 
other, and the non-uniqueness in such a case does not pose a serious problem.

Various results are shown in Table~\ref{Table 9}.
\begin{table}[hp]
\caption{Shift of the Bose--Einstein condensation temperature and analogous models}
\label{Table 9}       
\vskip 2mm
\centering
\begin{tabular}{cc}
\toprule
\textbf{Bose Condensate} & \textbf{Parameter $c_1$} \\ \midrule
Factor approximant, third order                 & 1.0248\\ \midrule 
Factor approximant, fourth order                 & 1.0959\\ \midrule
Borel with factors, second order      & 0.8165\\ \midrule
Borel-light with factors, $p$-optimal &1.2842\\ \midrule
Borel-light with factors, $u$-optimal &1.2722\\ \midrule
``Exact'' Monte Carlo                 & 1.3 $\pm$ 0.05\\ \midrule 
Fractional Borel-light with roots~\cite{fr}             & 1.2498 \\ \midrule
Modified Even Pad\'{e}~\cite{pa}       & \bf{1.2885} \\ \midrule  
Corrected  Pad\'{e}~\cite{cor}               & 1.386 \\ \midrule
Odd Pad\'{e}~\cite{pa}       & 0.985  \\ \bottomrule 
\end{tabular}
\end{table}
The best result is marked in bold in Table~\ref{Table 9}.

In the case of the Bose temperature shift, the Borel-light method, based on the idea of 
corrected Borel-type approximants, optimization, and correction with Pad\'{e} approximations 
works well. The Pad\'{e}  method modified for an even number of terms in the expansion 
gives a result that is close to the latter, even without optimization or Borel transformation. 
But, it should be noted here that the even approximation is close in spirit to the general 
idea of the corrected approximants, as described in the Section~\ref{comb} and in 
the paper~\cite{cor}.

Accurate results for the sought parameter are also found with optimal generalized 
Borel summation with iterated roots, $c_ 1=1.339$~\cite{bo}; with the optimal Mittag--Leffler 
summation with iterated roots with $c_1= 1.3397$~\cite{ml}; and with the corrected iterated 
roots, $c_1=1.3092$ (see~\cite{fr}, and references therein). Kastening 
\cite{Kas2004a, Kas2004b, Kas2004c}, using the optimized variational perturbation theory, 
estimated $c_1$ as $1.27 \pm 0.11$.

\subsection{Fermi Gas: Unitary Limit}
\label{uni}


The ground-state energy $E$ of a dilute Fermi gas can be obtained from perturbation theory
\cite{Bak1999}, so that 
\be
\label{7.14}
 E(g) \simeq a_0 + a_1 g + a_2 g^2 + a_3 g^3 + a_4 g^4 ,
\ee
with the coefficients
$$
 a_0 = \frac{3}{10} , \qquad a_1 = - \; \frac{1}{3\pi} , \qquad
a_2 = 0.055661 ,
$$
$$
a_3 = -0.00914 , \qquad a_4 = -0.018604 .
$$
The effective coupling parameter $g \equiv | k_F a_s | $ 
is simply related to the  Fermi wave number $k_F$ and the atomic scattering length $a_s$.  
The limit of very large effective coupling $g$ is called the unitary Fermi gas~\cite{Ket2008}. 
Monte Carlo simulations for the case of $g \ra \infty$~\cite{rev,num} yield
\be
\label{7.15}
 E(\infty) = 0.1116.
\ee
The experimental value \cite{rev,exper} $$ E(\infty) = 0.1128,$$ is rather close to the 
Monte Carlo results. 

Factor approximants in the third and fourth orders give rather high estimates for 
$E(\infty)$, as shown in Table~\ref{Table 10}, while Borel summation with 
factor approximants gives rather low estimates.

\begin{table}[hp]
\caption{Fermi gas energy in the unitary limit}
\label{Table 10}    
\vskip 2mm
\centering
\begin{tabular}{cc}
\toprule
\textbf{Fermi Gas} & \textbf{Unitary Limit}  \\ \midrule
Factor approximant, third order           & 0.174\\ \midrule
Factor approximant, fourth  order           & 0.1644\\ \midrule
Borel factors,  fourth  order         & 0.0898\\ \midrule
Borel-light  with factors, $p$-optimal, min.diff.            & \bf{0.1125} \\ \midrule
Borel-light  with factors, $p$-optimal, min.deriv.            & 0.1293\\ \midrule
``Exact''  Monte Carlo~\cite{rev,num}            & 0.1116\\ \midrule 
Borel-factor-light, Mittag--Leffler~\cite{ml}              & 0.1193  \\ \midrule
Generalized Borel-light with roots~\cite{fr}           & 0.1193 \\ \midrule
Fractional Borel-light with roots~\cite{fr}           & 0.1256  \\ \midrule
Borel with roots ~\cite{fr}           & 0.1329 \\ \midrule
Diagonal Pad\'{e}       &0.1705   \\ \bottomrule
\end{tabular}
\end{table}
The best result is marked in bold in Table~\ref{Table 10}.

None of the methods, including those using optimizations, bring accurate solutions. 
$u$-Optimization gives $B\approx 0.178$ with or without correction terms. The result 
\mbox{$B\approx 0.272375$} is found with $p$-optimization by means of the equation
$$
B_{3}\left(p\right) - B_{2}\left(p\right)=0,\, \, \, u=0,
$$
with the optimum at $p=p_2=-3.313127$. However, the Borel-light techniques give a holomorphic 
correcting approximant with the value of the amplitude in the sixth order being
$$
B=B_{3,2}^*(p_2) = P_{3/3}(\infty)\;  C_2(p_2)=0.11253.
$$
The corresponding approximant
$$
f_{3,2}^*\left(g\right)\simeq P_{3,3}\left(g\right) f_2^*\left(g\right)
$$
that leads to a very accurate estimate is shown below:
$$
f_{3,2}^*\left(g\right)=
\frac{0.0182215+0.226078 g+0.656425 g^2+0.123943 g^3}{0.0607385+0.753593 g+2.28882 g^2+g^3} 
\frac{\left(1+g\right)^{0.0438025}}{\left(1+9.07436 g\right)^{0.0438025}} \; .
$$

The lower-order approximants, $f_{1,2}^*(g), f_{2,2}^*(g),$ appear to be non-holomorphic. 

From $p$-optimization with the minimal derivative condition
$$
\frac{\partial B_{3}(p)}{\partial p}=0,
$$
we find $p=p_3=-3.1325$. The Borel-light technique then gives a reasonable estimate for 
the amplitude 
$$
B=B_{3,3}^*(p_3) = \; P_{3/3}(\infty)\;  C_3(p_3)=0.1293.
$$

We conclude that the Borel-light methods based on 
Borel summation with optimization and subsequent correction with Pad\'{e} approximations 
work well for the problem of a unitary limit, approaching the quality of Monte Carlo and 
experimental results.

\section{In Lieu of Conclusions}
\label{conc}

For the first time, Fractional Borel Summation was applied in conjunction with self-similar 
factor approximants. This is the main distinction from the method developed in~\cite{fr} 
in conjunction with the so-called self-similar iterated root approximants. It was found 
that the technique of Fractional Borel Summation can be most conveniently applied in hybrid 
form. Such hybrid 
approximants emerge when the Borel-transformed factor approximations are complemented 
multiplicatively with Pad\'{e} approximants to satisfy the original 
expansions asymptotically. A detailed comparison of different methods is performed on 
a large set of 
examples, including the approximation techniques of self-similarly modified 
Pad\'{e}-Borel approximations. Such a comparison clearly emphasizes the strong points 
of each of the techniques.

We confirm that 
the quality 
of the analytical reconstructions can match the quality of heavy numerical work. Analytical 
results are often very similar to the exact numerical data. They follow after a sequence 
of a few analytical steps, and the most difficult part of solving transcendental equations 
is performed numerically with any desired accuracy.

The discussed approach to the resummation of the asymptotic series is multi-leveled. 
General Borel Fractional transformation of the original series was introduced. The found 
transformed series should be resummed in order to adhere to the asymptotic power laws. 
One starts with the formulation of dynamics in the approximations space. To this end, 
self-similarity is employed. The flow in the approximation space should be controlled, 
and ``deep'' control is incorporated into the definitions of the self-similar approximants. 
Certain classes of self-similar approximants satisfying the power law behavior by design, 
such as root and factor self-similar approximants, are chosen for reasons of transparency, 
explicitness, and convenience. We also employ properly modified (taking account of power 
law behavior) Pad\'{e} approximants by noting that they can be viewed as a particular case 
of factor approximants. The asymptotic power law properties of the approximations follow 
directly from the ways in which controls are introduced into the approximation dynamics. 
The second level of controls concerns the dynamics of Borel-type approximations that emerge 
after an inverse transformation is accomplished back to the original space of approximations. 
Self-similarity of the approximants allows us to find the dependencies of the sought critical 
properties explicitly. The second level of control is then performed by applying the minimal 
difference and minimal derivative conditions with respect to the parameters that are explicit 
in the original Borel Fractional transformation. 

The main methods compared in the paper each have their own merits. Different methods, based 
on Pad\'{e} approximants~\cite{pa}, can be useful as benchmarks for the evaluation of the 
results. The standard scheme of odd Pad\'{e} approximants is not competitive with respect 
to the other considered methods. Of course, this statement concerns only the physical problems 
considered above. However, the methods based on Pad\'{e} approximants are indispensable for 
computations with very long expansions. For shorter expansions, the other methods discussed 
in the paper should be used. But, in many cases of such expansions, the diagonal Pad\'{e} 
approximants reappear in combination with factors and roots. Such hybrid forms of corrected 
approximants, when the approximations of different types are applied consequentially, are 
very useful for the practical purpose of accurate summation.


In conclusion, we recapitulate the main steps of the multi-level methodology developed 
in the present paper. 

\begin{enumerate}
\item 
The initial truncated series \eqref{2} is transformed into the form of a new, 
supposedly better behaving series. The chosen transformation is the Fractional Iterated 
Transform (see Section~\ref{frd}). At this stage, the control parameters are introduced. 
They have to be found at the final optimization step.
\item 
To arrive to the correct asymptotic behavior at infinity, the transformed series is 
approximated by the class of approximants with power law behavior \eqref{4} at infinity. 
At this step, we reconstruct the coefficients for an arbitrary $n$.
\item 
When the guiding principle of self-similarity, described in Section ref{self}, 
is applied together with the so-called algebraic transformation of the original 
series~\cite{Gluzman_28}, we arrive at two types of approximants with the desired 
property at infinity. These are the self-similar roots discussed in Section 
\ref{iter} and the self-similar factor approximants described in the Sections~\ref{fap} 
and~\ref{fact}. 
\item 
In the limit of a large $x$, the expressions for the critical amplitudes become 
explicit and factorize into the parts emerging from the self-similar approximants 
applied to the transformed series and from the $\Gamma$-functional terms emerging 
from the inverse transformation, being dependent on the type of transformation.
\item 
The application of self-similar iterated roots was considered previously~\cite{fr}.  
Now, after the optimization with the minimal difference and (or) minimal derivative 
conditions described in Sections~\ref{cf} and~\ref{frd}, we resort to the best 
known (to us) guiding principle for {\it{{numerical convergence of the sequences of the 
approximations}}} for the sought quantities. The uniqueness of the solution for the 
given class of approximants is achieved due to careful selection of the transformation 
to the original truncated~series.
\item 
In the current paper, we employ the self-similar factor approximants together with 
the most convenient technically and theoretically sound method of so-called hybrid 
techniques (see Sections~\ref{comb} and~\ref{light}). Typically, in low orders 
of the perturbation theory, an optimized factor approximant has to be found. To return 
to the original series, we restore the factors in the form of the diagonal Pad\'{e} 
approximants. Besides numerical convergence, we are also guided by the Gonchar 
results on the convergence of the diagonal Pad\'{e} approximants~\cite{gonch}. Such a 
selection leads, by design, to a unique limit that can be found numerically as the 
approximation for the sought quantity.
\end{enumerate}

The Fractional Borel Summation with iterated roots can be successfully applied for 
various problems~\cite{fr}, although it is not always the best method. Iterated roots 
can be exceptionally helpful for problems which appear to be indeterminate and those 
that are poorly treatable from the standpoint of the Pad\'{e} approximations~\cite{fr} 
and the factor approximations. The cases with fast-growing and rapidly diminishing by 
magnitude coefficients $a_n$ are better treated by means of Borel summations with 
iterated roots. The cases with slowly diminishing by magnitude coefficients $a_n$ are 
best approximated by Borel summations with factors. Fractional Borel Summation with 
factor approximants in its different realizations can be successfully applied to various 
problems of the types discussed above. In addition, it is shown to be the best method, 
or very close to it, for about half of the problems considered in the paper, much more 
often than the other methods discussed above. Thus, the techniques based on factors are 
most useful in the context of finding the best method for the particular problems. One 
can say that the factor approximants are a sharper tool than the iterated roots, but the 
iterated roots are useful for a wider range of applications.

\end{document}